\documentclass[12pt]{article}
\usepackage{epsf}
\usepackage{amsmath}
\usepackage{graphics}
\usepackage{cite}

\renewcommand{\thefootnote}{\#\arabic{footnote}}
\begin{document}

\newcommand{\gtrsim}{ \mathop{}_{\textstyle \sim}^{\textstyle >} }
\newcommand{\lesssim}{ \mathop{}_{\textstyle \sim}^{\textstyle <} }

\newcommand{\rem}[1]{{\bf #1}}

\renewcommand{\thefootnote}{\fnsymbol{footnote}}
\setcounter{footnote}{0}
\begin{titlepage}

\def\thefootnote{\fnsymbol{footnote}}

\begin{center}
\hfill hep-th/0511265\\
\hfill December 2005\\
\vskip .5in
\bigskip
\bigskip
{\Large \bf Suggestion of Explicit Field Transformations underlying Misaligned Supersymmetry}

\vskip .45in

{\bf Paul H. Frampton}

{\em Perimeter Institute, 31 Caroline Street, Waterloo, ON N2L 2Y5, Canada}\\

and

{\em University of North Carolina, Chapel Hill, NC 27599-3255, USA \footnote{Permanent address}}

\end{center}

\vskip .4in
\begin{abstract}
In order that nonsupersymmetric quiver gauge theories can satisfy naturalness
requirements to all orders of perturbation theory, one expects 
a global symmetry similar to, but different from, supersymmetry. Consistent
with the generalized no-go theorem published by Haag {\it et al} in 1975, we suggest 
a generalization of supersymmetry to a misaligned supersymmetry where
fermionic generators do not commute with gauge transformations.
An explicit form for the corresponding field transformations is suggested.
\end{abstract}
\end{titlepage}

\renewcommand{\thepage}{\arabic{page}}
\setcounter{page}{1}
\renewcommand{\thefootnote}{\#\arabic{footnote}}
\newpage

\bigskip

\noindent {\it I. Introduction}

\bigskip

It is an old notion that ${\cal N}=4$ supersymmetric gauge theory
is germane to the generalization of the standard model of particle
phenomenology. The ${\cal N}=4$ theory has remarkable properties
which include ultra violet finiteness and conformal invariance.

Nevertheless, one striking feature of the standard model is the
presence of chiral fermions which excludes both ${\cal N}=4$ and
${\cal N}=2$ extended supersymmetries. Also, although the situation
might change, the absence of any experimental
support even for ${\cal N}=1$ supersymmetry is striking.

This has led to reconsideration of the nonsupersymmetric
${\cal N}=0$ case, including ${\cal N}=0$
quiver gauge theories with an $U(N)^n$ gauge group and
matter fields in bifundamental representations.
Here one obstacle is a theoretical one in the form
of well-known no-go theorems.

Before the discovery of supersymmetry, Coleman and Mandula\cite{CM}
proved a no-go theorem published in 1967 which stated that
the only possible symmetries of the S matrix were products
of spacetime and internal symmetries. Shortly after the discovery
of supersymmetric theories, Haag, Lopuszanski and Sohnius
generalized this to show in a 1975 publication \cite{HLS}
that under certain assumptions, supersymmetry is the only possibility. 

No-go theorems can be useful because
they provide a set of assumptions some or all of which must be violated in order
to make progress. For example, Coleman and Mandula considered
only Lie algebras with commutators and not graded algebras
with anticommutators. Here, I shall suggest that the way around 
the generalized no-go theorem of Haag, Lopuszanki and Sohnius is to 
relax their assumption that
the generators of the symmetry commute with gauge transformations.
In particular, for a $U(N)^n$ quiver gauge theory, we suggest
fermionic generators which transform as bi-bifundamentals under $U(N)^n$.

In Section II, the symmetry transformations of ${\cal N}=4$
Yang-Mills are briefly reviewed.
In Section III, the explicit form of a misaligned supersymmetry
transformation is suggested, with fermionic generators which transform
under the gauge group. Finally in Section IV there is
discussion.

\newpage

\bigskip

\noindent {\it II. ${\cal N}=4$ supersymmetry}

\bigskip

\noindent Here we collect briefly some well-known facts, for convenience.

\bigskip

\noindent The action for ${\cal N}=4$ Yang-Mills can be written
\begin{eqnarray}
S & = &
\int d^4x \left[ -\frac{1}{4} F_{\mu\nu a}F^{\mu\nu a}
+ \frac{1}{2} D_{\mu} \Phi_{ij}^{a} D^{\mu} \Phi_{ij}^{a}
+i \bar{\chi}^{a} \gamma.D L \chi^{a} \right.  \nonumber \\
&  &
\left.   - \frac{1}{2} i g f_{abc} \left( \bar{\tilde{\chi}}^{ai} L \chi^{jb}
\Phi_{ij}^{c}
- \bar{\chi}_{i}^{a} R \tilde{\chi}_{j}^{b} \Phi^{ijc} \right)  \right.  \nonumber \\
& & \left.  - \frac{1}{4} g^2 \left(
f_{abc} \Phi_{ij}^{b} \Phi_{kl}^{c} \right)
\left(
f_{ade} \Phi^{ijd} \Phi^{kle} \right)
\right]
\label{N=4S}
\end{eqnarray}
where $\mu,\nu = 0,  1,  2,  3$; $i, j, k, l = 1, 2,  3$; $L=\frac{1}{2} (1+\gamma_5)$,
$R=\frac{1}{2}(1-\gamma_5)$; and $\tilde{\chi}_i = C\bar{\chi}^{iT}$ with $C$ the charge
conjugation operator. 

\bigskip

\noindent The action (\ref{N=4S}) is invariant under the ${\cal N}=4$ supersymmetry
\cite{BSS}

\begin{eqnarray}
\delta A_{\mu}^{a} & =  & i \left(\bar{\alpha}_i \gamma_{\mu} L \chi^{ia}
- \bar{\chi}_i^a \gamma_{\mu} L \alpha^i \right).  \nonumber  \\
\delta \Phi_{ij}^{a} & = &  i \left( \bar{\alpha}_j R \tilde{\chi}_i^a
- \bar{\alpha}_i R \tilde{\chi}_i^a
+ \epsilon_{ijkl} \bar{\alpha}^k L \chi^{la}  \right).   \nonumber  \\
\delta L \chi^{ia} & = & 
\sigma_{\mu\nu} F^{\mu\nu a} L \alpha^i
- \gamma.D \Phi^{ij a} R \tilde{\alpha}_j
+ \frac{1}{2} g f_{abc} \phi_b^{ik} \Phi_{kj}^{c} L \alpha^{j}   \nonumber   \\
\delta R \tilde{\chi}_i^a & = & 
\sigma_{\mu\nu} F^{\mu\nu a} R \tilde{\alpha}_i
+ \gamma.D \Phi_{ij}^a L \alpha^i 
+ \frac{1}{2} g f_{abc} \Phi^b_{ik} \Phi^{kj}_{c} L \tilde{\alpha}_{j}.   
\label{susy}
\end{eqnarray}
where $\alpha^i$ transforms as a ${\bf 4}$ and $\bar{\alpha}_i$
as a ${\bf \bar{4}}$ under an internal $SU(4)$ symmetry.

\bigskip

\noindent The group indices $a,b,c$ run over the dimension of the gauge group
$a, b, c = 1,.....,d_G$. For $G = SU(N)~~ {\rm or} ~~ U(N)$, $d_G = (N^2-1) ~~ {\rm or} ~~
N^2$ respectively. Note that the infintesimal supersymmetry parameter $\alpha^i$
is singlet under the gauge group $G$. This assumption will be
relaxed for misaligned supersymmetry in the next section.

\bigskip

\newpage

\bigskip

\noindent {\it III. Misaligned supersymmetric gauge field theory (MSGFT)}

\bigskip

The name is taken from \cite{Dienes}
where string models without supersymmetry were studied, particularly
the supertrace conditions necessary for cancellation of
ultra violet divergences.  The nonsupersymmetric quiver gauge 
theories introduced
in \cite{PHF1998} and analyzed further in \cite{catchall}
satisfy such supertrace conditions 
if all scalars are in bifundamentals\cite{CFR}
so the name ``misaligned" supersymmetric gauge
field theory (MSGFT)
is appropriate.
In \cite{Dienes}, however, no explicit field transformation
underlying misaligned supersymmetry was given 
and my aim here is to suggest how this may be
accomplished.

\bigskip

More recently, in \cite{CDF}, it was discussed how chiral trangle
anomalies can be compensated in MSGFT. 

\bigskip

A specific MSGFT model is defined by several integers, namely $N$ (the number of coalescing
parallel D3 branes in AdS/CFT, also the $N$ in the gauge group
$U(N)^n$), $n$ (defining the abelian orbifold group $Z_n$,
also the $n$ in the gauge group $U(N)^n$); and 
three integers $A_1, A_2, A_3$ which specify the embedding
$Z_n \subset SU(4)$ where $SU(4)$ is the internal symmetry of the ${\cal N}=4$
case corresponding to replacing the orbifold by a manifold.
Note that the fourth integer $A_4$ defining the transformation of the {\bf 4}
of $SU(4)$ is not independent because $A_4 = - A_1 - A_2 - A_3$ (mod $n$).
In summary, MSGFT  models (of the subclass studied in
\cite{PHF1998,catchall}) are specified by five integers $\{N, n, A_1, A_2, A_3\}$.

\bigskip

The action for such a MSGFT in the present notation
(adapted from \cite{BSS,CFR}) is
\begin{eqnarray}
S & = &
\int d^4x \left[ -\frac{1}{4} F_{\mu\nu a;r,r}F_{r,r}^{\mu\nu a}
+ \frac{1}{2} D_{\mu} \Phi_{ij;r+a_i,r}^{a} D^{\mu} \Phi_{ij;r,r+a_i}^{a}
+i \bar{\chi}_{r+A_m,r}^{a} \gamma.D L \chi_{r,r+A_m}^{a} \right.  \nonumber \\
&  &
\left.   - \frac{1}{2} i g f_{abc} \left( \bar{\tilde{\chi}}_{r,r+A_m}^{ai} L \chi_{r+A_m,r+A_m+A_n}^{jb}
\Phi_{ij;r+A_m+A_n,r}^{c} \right. \right. \nonumber  \\
&  & - \left.  \left. \bar{\chi}_{i;r,r+A_m}^{a} R \tilde{\chi}_{j;r+A_m,r-A_n}^{b} \Phi_{r-A_n,r}^{ijc} 
\right)  \right.  \nonumber \\
& & \left.  - \frac{1}{4} g^2 \left(
f_{abc} \Phi_{ij;r,r+a_i}^{b} \Phi_{kl;r+a_i,r+a_i+a_j}^{c} \right)
\left(
f_{ade} \Phi_{r+a_i+a_j,r+a_j}^{ijd} \Phi_{r+a_j,r}^{kle} \right)
\right]
\label{MSGFTaction}
\end{eqnarray}
in which the $a_i$ are defined by $a_i = A_2+A_3, a_2=A_3+A_1, a_3=A_1+A_2$;
the subscript $r=1, 2, .... n$ is a node label; when the two node 
superscripts are equal it is an adjoint plus
singlet of that $U(N)_r$; when the two subscripts are unequal it is a bifundamental
and the two gauge labels transform under different $U(N)$ gauge groups.

\newpage

\bigskip

Now we address the question of what variation of the
fields in the action (\ref{MSGFTaction}) will leave it invariant.
Given the field content, the infinitesimal fermionic
parameters must transform under the $U(N)^n$ gauge group.
As a generalization of equations (\ref{susy}), we suggest
\begin{eqnarray}
\delta \left( A^{(p)}_{\mu} \right)^{\beta_p}_{\alpha_p} & =  & i \left( [\bar{\alpha}_i]^{\beta_p,\gamma_p}
_{\alpha_p,\delta_{p+A_m}} \gamma_{\mu} L \left( \chi^{i(p,p+A_m)} \right)^{\delta_{p+A_m}}_{\gamma_p}
- \left( \bar{\chi}_i^{(p-A_m,p)} \right)^{\delta_p}_{\gamma_{p-A_m}} \gamma_{\mu} L 
[\alpha^i]^{\alpha_p,\gamma_{p-A_m}}_{\alpha_p,\delta_p} \right) \nonumber  \\
\delta \left( \Phi_{ij}^{(p,p+a_i)}\right)^{\alpha_{p+a_i}}_{\alpha_p} & = &  
i \left( [\bar{\alpha}_j]^{\alpha_{p+a_i},\beta_{p}}_{\alpha_p,\beta_{p+A_m}} R
\left( \tilde{\chi}_i^{(p,p+A_m)} \right)_{\beta_{p}}^{\beta_{p+A_m}}
- [\bar{\alpha}_i]^{\alpha_{p+a_i},\beta_{p}}_{\alpha_p,\beta_{p+A_m}} R
\left(\tilde{\chi}_j^{(p,p+A_m)} \right)_{\beta_{p}}^{\beta_{p+A_m}}  \right. \nonumber \\
& & \left. + \epsilon_{ijkl} [\bar{\alpha}^k]^{\alpha_{p+a_i},\beta_p}_{\alpha_p,\beta_{p+A_m}} L 
\left( \chi^{l (p,p+A_m)} \right)_{\beta_p}^{\beta_{p+A_m}}  \right). \nonumber  \\
\delta \left( L \chi^{i(p,p+A_m)} \right)_{\alpha_p}^{\alpha_{p+A_m}} & = & 
\sigma_{\mu\nu} \left( F^{\mu\nu (p)} \right) _{\beta_p}^{\gamma_p}  
L [\alpha^i]_{\alpha_p \gamma_p}^{\alpha_{p+A_m} \beta_p} \nonumber \\
& & - \gamma.D \left( \Phi^{ij (p,p+a_i)} \right)_{\beta_p}^{\beta_{p+a_i}}  R 
[\tilde{\alpha}_j]_{\alpha_p \beta_{p+a_i}}^{\alpha_{p+A_m} \beta_p}  \nonumber  \\
& & + \frac{1}{2} g \epsilon_{\alpha_{p+a_i}\beta_{p+a_i}\gamma_{p+a_i}}
\epsilon^{\beta_p\gamma_p\delta_p}
\left( \phi^{ik (p,p+a_i)} \right)_{\gamma_p}^{\beta_{p+a_i}}
\left( \Phi_{kj}^{(p,p+a_i)} \right)_{\delta_p}^{\gamma_{p+a_i}}
L [\alpha^{j}]_{\alpha_p\beta_p}^{\alpha_{p+A_m}\alpha_{p+a_i}} \nonumber \\ 
\delta \left( R \tilde{\chi}_{i}^{(p-A_m,p)} \right)_{\alpha_{p-A_m}}^{\alpha_{p}} & = & 
\sigma_{\mu\nu} \left( F^{\mu\nu (p)} \right) _{\beta_p}^{\gamma_p}  
R [\tilde{\alpha_i}]_{\alpha_{p-A_m} \gamma_p}^{\alpha_{p} \beta_p} \nonumber \\
& & + \gamma.D \left( \Phi_{ij}^{(p-a_i,p)} \right)_{\beta_{p-a_i}}^{\beta_{p}}  L 
[\alpha^j]_{\alpha_{p-A_m} \beta_{p}}^{\alpha_{p} \beta_{p-a_i}}  \nonumber  \\
& & + \frac{1}{2} g \epsilon_{\alpha_{p+a_i}\beta_{p+a_i}\gamma_{p+a_i}}
\epsilon^{\beta_p\gamma_p\delta_p}
\left( \phi_{ik}^{(p,p+a_i)} \right)_{\gamma_p}^{\beta_{p+a_i}}
\left( \Phi^{kj (p,p+a_i)} \right)_{\delta_p}^{\gamma_{p+a_i}}
R [\tilde{\alpha_{j}}]_{\alpha_{p-A_m}\beta_p}^{\alpha_{p}\alpha_{p+a_i}} 
\label{misalignedsusy}
\end{eqnarray}

\bigskip

\noindent The equations (\ref{misalignedsusy}) are written so that they reduce
to the ${\cal N}=4$ equations (\ref{susy}) when the internal $U(N)^n$ dependence of the fermionic
generators is removed and are written such that the transformation properties
under the gauge group $U(N)^n$ are consistent for each term in the
field transformations (\ref{misalignedsusy}). 

In the limit $A_m=a_i=0$ and $n=1$, the bifundamentals become adjoints and
the couplings in the transformations Eq.(\ref{misalignedsusy})
reduce to those in Eq.(\ref{susy}); this requirement excludes 
further (symmetric) cubic couplings in Eq. (\ref{misalignedsusy}).

We see that the infinitesimal generators n Eq. (\ref{misalignedsusy})
must generically be outer products
of two bifundamentals under $U(N)^n$ although in all 
terms of (\ref{misalignedsusy}) this reduces
to an outer product of one adjoint with one bifundamental.
In the transformation of the $\chi^i$ fields 
I have for definiteness specialized
to the case $N=3$ in generalizing the structure constants $f_{abc}$
of (\ref{susy}) for adjoint representations to the antisymmetric tensors 
$\epsilon_{\alpha\beta\gamma}$ in (\ref{misalignedsusy}) for
bifundamental representations; for  
general $N$ one can form
\footnote{I thank Professor T.W. Kephart for this remark.}
a unique antisymmetric cubic invariant from bifundamentals writable in two equivalent forms
\begin{equation}
f_{abc} (\lambda^a)^i_{i^{'}} (\lambda^b)^j_{j^{'}} (\lambda^c)^k_{k^{'}}
\Phi^{i^{'}}_i \Phi^{j^{'}}_j \Phi^{k^{'}}_k ~~~ {\rm or} ~~~ 
\epsilon^{ijklmn...xyz}\epsilon_{i'j'k'lmn...xyz} \Phi_i^{i^{'}} \Phi_j^{j^{'}}
\Phi_k^{k^{'}}
\end{equation}

\newpage

\bigskip

\noindent {\it IV. Discussion.}

\bigskip

A first issue concerns the no-go theorems of
\cite{CM,HLS}. There is no problem with \cite{CM}
which did not consider fermionic generators and the
generalized no-go theorem in \cite{HLS} implicitly 
assumes that the fermionic generators are singlets
under the gauge group; since this assumption is
violated in misaligned supersymmetry, the no-go
theorem \cite{HLS} is inapplicable.

\bigskip

There remain a number of questions to be explored:
Does variation under the field transformations (\ref{misalignedsusy})
really provide an exact symmetry of the action (\ref{MSGFTaction})? 
Do the generators form a closed algebra and the transformations
a group?
What are the representations of this group? The quiver diagram must
form a representation but it may be reducible. It would be interesting to
know the irreducible representations.
Do MSGFT share properties of supersymmetric gauge theories
such as non renormalization theorems?
Can a MSGFT be conformally invariant?

\bigskip

It opens the door for research to study TeV scale 
conformality models, alternative to TeV scale supersymmetry.
Experiment will enable us to ascertain the approach favored
by Nature.

\bigskip
\bigskip
\bigskip

\begin{center}

{\bf Acknowledgements}

\end{center}

\bigskip

The hospitality of the Perimeter Institute is acknowledged. This research was
supported in part by the U.S. Department of Energy under Grant No.
DE-FG02-97ER-41036.

\end{document}